\documentclass[twocolumn,amsmath,amssymb]{snp}
\usepackage{graphicx}% Include figure files
\usepackage{dcolumn}% Align table columns on decimal point
\usepackage{bm}% bold math
\newcommand{\nn}{\nonumber}
\newcommand{\be}{\begin{equation}}
\newcommand{\ee}{\end{equation}}
\newcommand{\bea}{\begin{eqnarray}}
\newcommand{\eea}{\end{eqnarray}}

\begin{document}

\title{{\Large Collisional time and shear relaxation time for interacting QGP}}% Force line breaks with \\

\author{\large Ankit Anand$^{1,*}$, Souvik Paul$^{1}$, Sarthak Satapathy$^2$,
Sabyasachi Ghosh$^2$}
\email{ankit.anand0123@gmail.com}
\affiliation{$^1$Department of Physical Sciences, Indian Institute of Science Education 
and Research Kolkata, Mohanpur, West Bengal 741246, India}
\affiliation{$^2$Indian Institute of Technology Bhilai, GEC Campus, Sejbahar, Raipur 492015, 
Chhattisgarh, India}
\maketitle

We have estimated two different time scales of interacting quark gluon plasma (QGP)
system. One is collisional time scale $\tau_c$, which carry microscopic interaction of quarks
and gluons in medium. Another is shear relaxation time $\tau$, tuning the strength of shear 
viscosity coefficients QGP. Following the comments of Ref.~\cite{Muronga}, two time scales
are different but no general formula linking $\tau$ and
$\tau_c$ exists; their relationship depends in each case on the system under consideration.
The present work has attempted to estimate these two time scales for interacting QGP 
and to realize their comparative strengths.  

Let us first estimate the $\tau_c=1/\Gamma_c$, where $\Gamma_c$ is thermal width
of quarks or gluons. We can understand the transition from non-interacting to interacting
picture of QGP system as transition from $\Gamma_c=0$ to $\Gamma_c\neq 0$. 
It provides a possibility to map interaction of QGP by introducing
a finite thermal width of quarks and gluons. This well-standard technique
 can be found in Ref.~\cite{Cassing} and references therein.
 \begin{figure}
\centering
\includegraphics[scale=0.27]{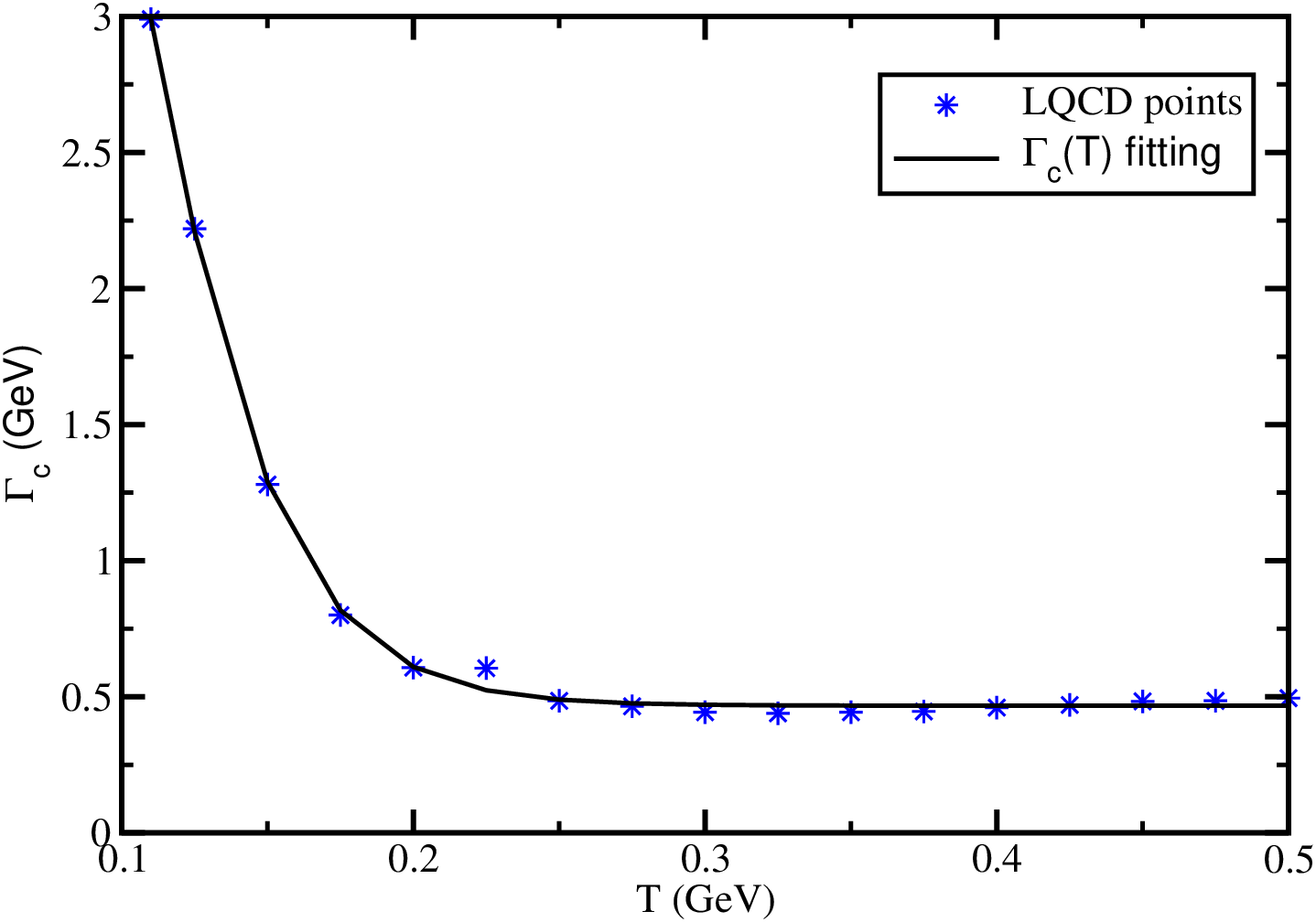}
\includegraphics[scale=0.27]{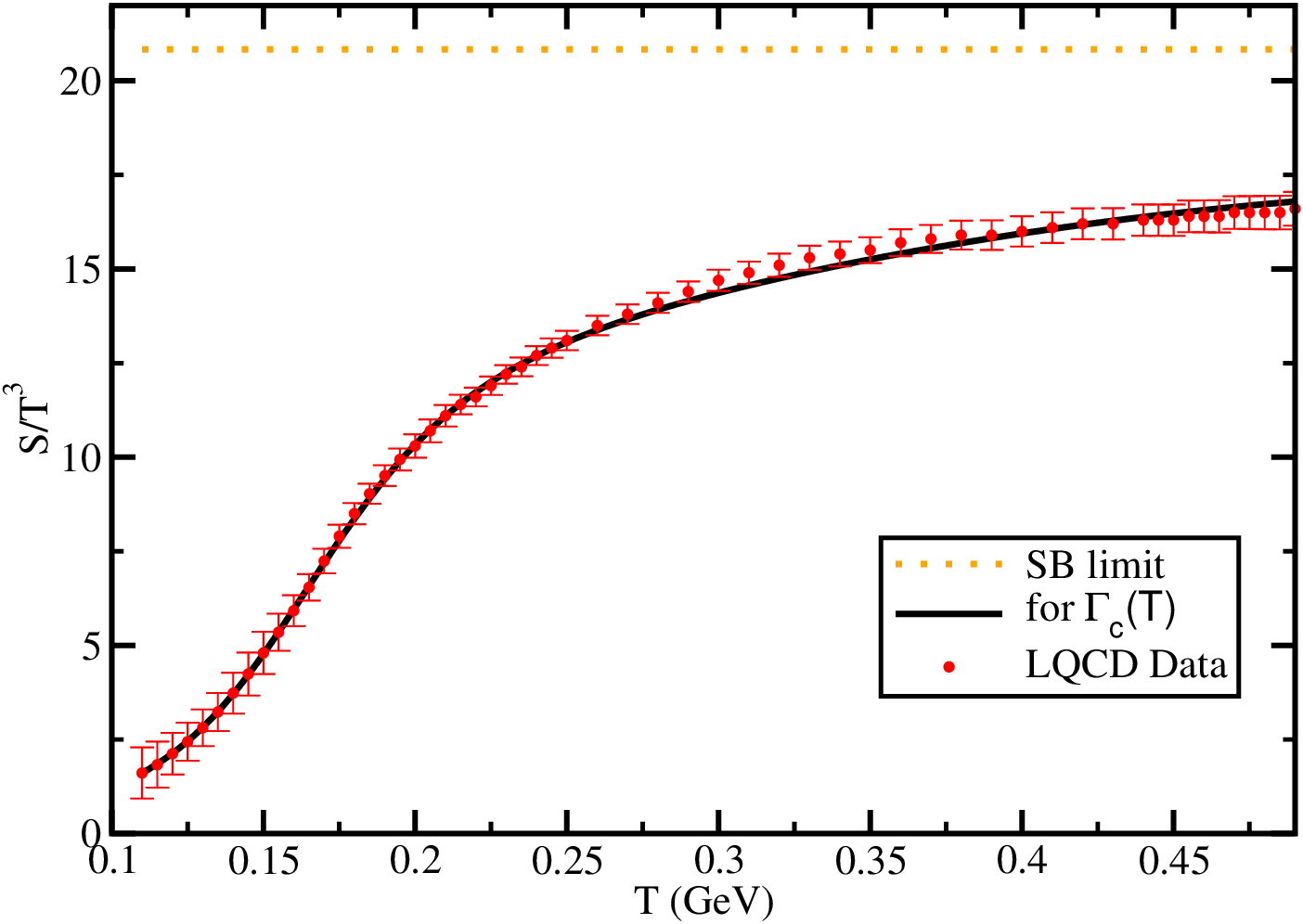} 
\caption{(a) Temperature dependence thermal width $\Gamma_c(T)$ parametrization curve (solid line)
	and LQCD extracted points (stars). (b) Their corresponding $s/T^3$ plots, where
	straight horizontal dotted line indicates SB limits of $s/T^3$.}
\label{GammavsT}
\end{figure}

Using spectral function $\rho(M, m_i)$ of quarks and gluons for finite $\Gamma_c$ in
interacting picture, we can express entropy density as
\bea
s &=& \sum_{i=u,d,s,g}g_i\int_0^\infty dM\rho(M, m_i)
\int_{0}^{\infty}\frac{d^3p}{(2\pi)^3}
\nn\\
&&\Big(\sqrt{p^2+M^2}+\frac{p^2}{3\sqrt{p^2+M^2}}\Big)
\nn\\
&&\frac {1}{{\rm exp}(\beta \sqrt{p^2+M^2})-a_i}~,
% \nn\\
% &+& g_u\int_0^\infty dM\delta(M-m_u)
% \int_{0}^{\infty}\frac{d^3p}{(2\pi)^3}\frac{\sqrt{p^2+M^2}}{{\rm exp}(\beta \sqrt{p^2+M^2})+1}
% \nn\\
% &+& g_s\int_0^\infty dM\delta(M-m_s)
% \int_{0}^{\infty}\frac{d^3p}{(2\pi)^3}\frac{\sqrt{p^2+M^2}}{{\rm exp}(\beta \sqrt{p^2+M^2})+1}~,
\label{s_QGP_Gm}
\eea
where $a_i=\pm 1$ for fermion ($u$, $d$, $s$ quarks), boson (gluon $g$) respectively, and
the spectral functions of constituents of non-interacting and interacting medium are respectively
\bea
\rho(M, m_i)&=&\delta(M-m_i)
\nn\\
\rho(M, m_i)&=&\frac{1}{\pi}\Big(\frac{\Gamma_c}{\Gamma_c^2+(M-m_i)^2}\Big)~,
\eea
where $M$, $m_i$ is off-shell, on-shell mass of constituents. 
One can get back delta distribution
for vanishing thermal width because of relation
\be
\delta(M-M_i)=\lim_{\Gamma_c\rightarrow 0}\rho(M, m_i)~.
\label{delta}
\ee
Hence, in non-interacting picture ($\Gamma_c\rightarrow 0$) as well as massless limit,
Eq.~(\ref{s_QGP_Gm}) merges to standard Stephan-Boltzmann (SB) limiting value:
\be
s=\Big[g_g+(g_u+g_s)\Big(\frac{7}{8}\Big)\Big]\frac{4\pi^2}{90}T^3\approx 20.8~T^3~.
\label{s_QGP_m0}
\ee
According to lattice Quantum Chromo Dynamics (LQCD) calculation~\cite{LQCD1}, 
the numerical values of $s$ for QGP remain always lower than its SB limits. It can be
seen from Fig.~\ref{GammavsT}(b). By tuning $\Gamma_c$ in Eq.~(\ref{s_QGP_Gm}), we have
match LQCD data~\cite{LQCD1}, where we get parametrized of $\Gamma_c(T)$:
\be
\Gamma_c(T)=a_0-\frac{a_1}{e^{a_2(T-a_3)}+ a_4}
\label{Gm_fit}
\ee
with $a_0=6.76802$, $a_1=88.6265$, $a_2=-37.3715$, $a_3=0.170$, $a_4=14.0653$.
It is plotted in Fig.~\ref{GammavsT}(a).
% In hadronic temperature range, $\Gamma_c(T)$ decreases with $T$ but it
% saturates with the values $\Gamma_c\approx 0.500$ GeV in quark temperature domain.
% Here, one can again find set-2 parametrization of $\Gamma_c(T)$, where $\Gamma_c\rightarrow 0$
% at $T\rightarrow\infty$ but that choice provide a very bad matching of LQCD data, so we
% have not considered that set. 

%
\begin{figure}
	\centering
	\includegraphics[scale=0.27]{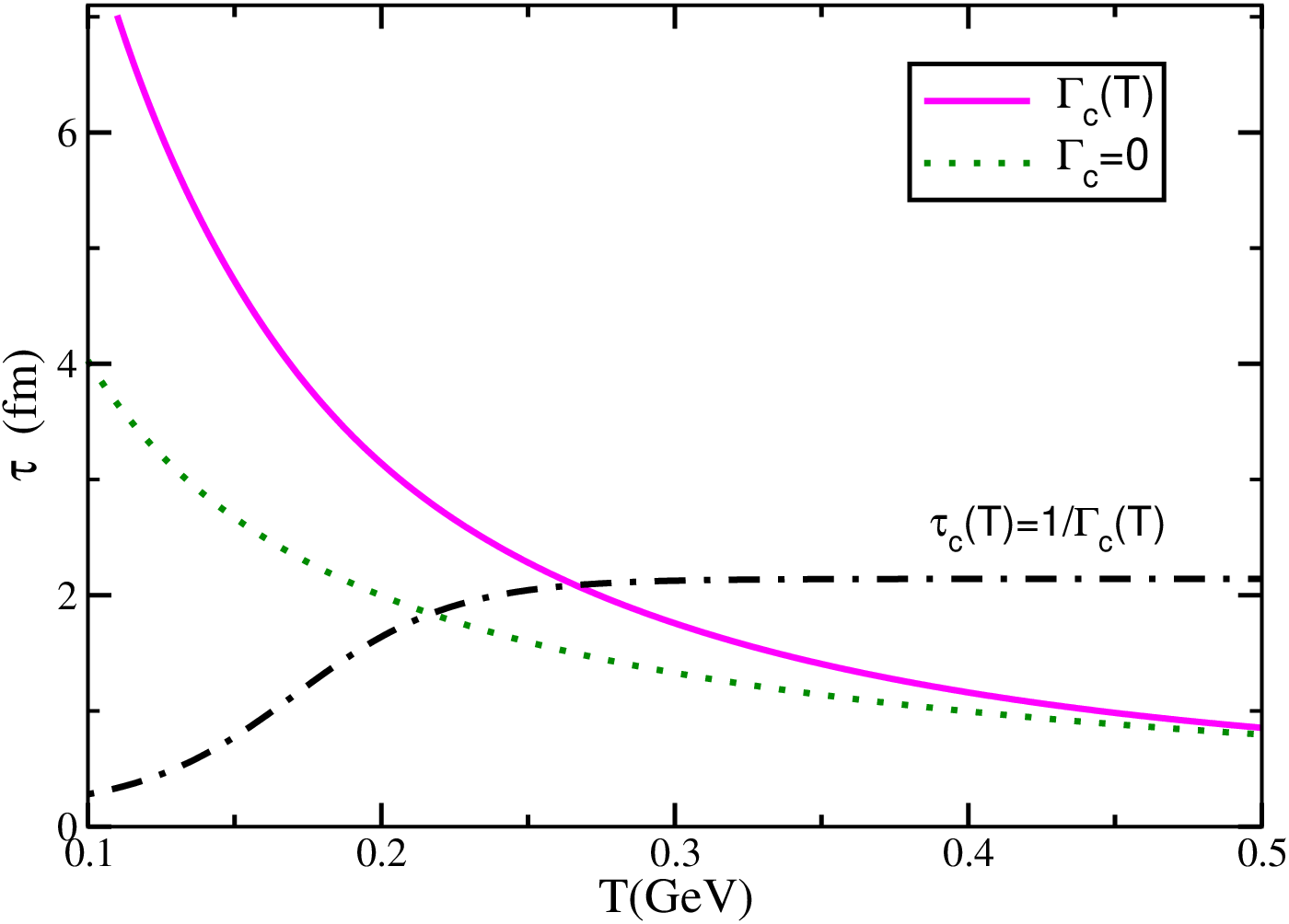} 
	\caption{By imposing
	$\eta/s=1/(4\pi)$, $\tau(T)$ has been found for non-interacting or $\Gamma_c=0$ (dotted line) and interacting or
        $\Gamma_c(T)$ (solid line) cases. The $\tau_c(T)=1/\Gamma_c(T)$ is also plotted to compare 
        with relaxation time scale $\tau$.}
	\label{EtaStauC}
\end{figure}
Now let us focus on other time scale - shear relaxation time $\tau$,
which basically tune the shear viscosity expression
\bea
\eta &=& \tau\sum_{i=u,d,s,g}g_i\int_0^\infty dM\rho(M, m_i)
\int_{0}^{\infty}\frac{d^3p}{(2\pi)^3}
\nn\\
&&\Big(\frac{p^4}{p^2+M^2}\Big)
\frac {{\rm exp}(\beta \sqrt{p^2+M^2})}{{\rm exp}(\beta \sqrt{p^2+M^2})-a_i}~,
\label{eta_tau}
\eea
The $\tau$ in $\eta$ can be guessed from experimental data of QGP fluid, which indicates about its perfect fluid
nature i.e. $\eta/s$ touch the KSS value $1/(4\pi)$. So
imposing $\eta/s=1/(4\pi)$ for non-interacting and interacting picture, based on $\Gamma_c(T)$ parametrization,
we have generated dotted and solid lines respectively in Fig.~\ref{EtaStauC}. Now,
drawing collisional time $\tau_c=1/\Gamma_c$ by dash-dotted line in Fig.~\ref{EtaStauC},
we can get a pictorial knowledge of $\tau$ and $\tau_c$ in one frame.
% 
% Let us try to relate these two time scale
% roughly as $\tau=\phi(T)\tau_c$, where $\phi(T)< 1$ for quark temperature domain and
% $\phi(T)> 1$ for hadronic temperature domain are noticed in Fig.~\ref{EtaStauC}. 
% If we roughly understand larger time scale
% as more macroscopic, then at quark temperature domain, $\tau_c$ is appeared as macroscopic
% scale, whereas at hadronic temperature domain $\tau$ plays the macroscopic role. 
In kinetic theory approximation, we generally consider $\tau\approx\tau_c$, which might be
more or less applicable near and above transition temperature at least in order of 
magnitude ($\tau_c\approx2$ fm, $\tau\approx 1$ fm). It indicates that high temperature 
QCD interaction time scale, covered by LQCD data is quite well agreement with shear dissipative 
interaction of QGP. A detail investigation on it can be seen in Ref.~\cite{LQCD_QGP}.

%\newpage
\end{document}